\documentclass[11pt,a4paper,twoside]{article}
\usepackage{natbib}
\usepackage{color}
\usepackage{graphicx}
\usepackage[affil-it]{authblk}
\usepackage{amssymb}
\usepackage{amsmath}

\begin{document}

\title{Estimating coupling directions in the cardio-respiratory system using recurrence properties}
\author[1]{Norbert Marwan}
\author[2,3,1]{Yong Zou}
\author[4]{Niels Wessel}
\author[4]{Maik Riedl}
\author[1,4]{J\"urgen Kurths}

\affil[1]{Potsdam Institute for Climate Impact Research, 14412 Potsdam, Germany}
\affil[2]{Department of Physics, East China Normal University, 200062, Shanghai, China}
\affil[3]{Department of Electronic and Information Engineering, The Hong Kong Polytechnic University, Hung Hom, Kowloon, Hong Kong}
\affil[4]{Institute of Physics, Humboldt-Universit\"at zu Berlin, Robert-Koch-Platz 4, 10099 Berlin, Germany}

\date{}
\maketitle

\abstract{The asymmetry of coupling between complex systems can be studied by conditional
probabilities of recurrence, which can be estimated by joint recurrence plots.
This approach is applied for the first time on experimental data: time series of the human cardio-respiratory system
in order to investigate the couplings between heart rate, mean arterial
blood pressure, and respiration. We find that the respiratory system couples towards
the heart rate and the heart rate towards the mean arterial blood pressure. 
However, our analysis could not detect a clear coupling direction between
the mean arterial blood pressure and respiration.}

\section{Introduction}
The cardio-respiratory system is complex with direct and indirect interactions in its sub-components. 
It includes not only mechanical components reflecting the changing pressure in the thoracic region but also the autonomic nervous control of both systems as well as a control of diaphragm and external intercostal muscles by means of the somatic nervous system.
Investigating and understanding of the couplings can help to identify and characterize different physiological and even pathological states, important for diagnosing and assessment of diseases \citep{hoyer2002,peupelmann2009,baer2010}.

Different linear and nonlinear approaches have been applied for studying couplings within the cardio-respiratory system, like spectral analysis \citep{eckberg2009point,faes2004causal,faes2010extended,taylor1996fundamental}, Granger causality \citep{faes2008assessment,riedl2010}, phase dynamics \citep{schaefer1998,porta2011causal}, conditional information \citep{faes2011information,faes2012non}, joint symbolic dynamics \citep{suhrbier2010cardiovascular,kabir2011}, \textcolor{black}{and model-based linear closed loop approaches \citep{baselli1994,porta2012}}. 
The main findings are a dependence of the couplings from the body position where the interaction between respiration and heart rate is dominant during supine position. 
In contrast to that the connection between heart rate and systolic blood pressure dominates upright position. In all these approaches, the considered data has been on a beat-to-beat basis. 
There are only few works which use continuous signals \citep{milde2011time,mullen1997system}. 
The most important difference is the use of a continuous blood pressure with its pulsating characteristic. 
Systolic and diastolic pressure corresponds only to events in this cyclic change. 
Therefore, an extension to a continuous signal leads to a difficult interpretation. 
For this reason the mean blood pressure value is considered which does not include the pulse pressure (systolic pressure-diastolic pressure) variability. 
This causes a reduced high frequency component of the mean blood pressure which corresponds to the mechanical influence of respiration.

Recently, a new non-linear method for studying coupling directions have been proposed, which is based on recurrence plots \citep{romano2007, hirata2010,zou2011}. 
A recurrence plot (RP) itself is a powerful concept allowing the investigation of a variety of aspects of complex systems, like transition studies, dynamical regime characterization, synchronization analysis, or surrogate constructions \citep{marwan2007,marwan2008epjst}. 
Bivariate extensions are cross RPs and joint RPs, which can be used to study complete and generalized synchronization, respectively \citep{marwan2007}. A further RP based approach to study couplings between two systems is based on the {\it probability of recurrence} and can be used to detect phase synchronization \citep{marwan2007}. 
This latter approach can be extended to a conditioned version allowing to infer coupling directions \citep{romano2007}. 
Its potential has been demonstrated on prototypical model systems, but not yet on experimental data. 

Here we will apply the approach of conditioned mean recurrence probabilities for the first time on experimental data which come from a study on cardiovascular variability in pregnancy and their change during preeclampsia. In general, the heart rate and the mean blood pressure increase with pregnancy \citep{eneroth1994autonomic}. A change of the coupling between heart rate and systolic blood pressure may also be observed \citep{baumert2002joint}. By definition, the blood pressure is significantly larger in preeclampsia than in normal pregnancy. But there is also a change in cardiovascular regulation which is indicated by decreasing respiratory sinus-arrhythmia \citep{eneroth1994autonomic} as well as an increased respiratory influence on diastolic blood pressure and a higher number of baroreflex events, an influence of systolic blood pressure on heart rate \citep{malberg2007}. The changes of the respiratory influence on heart rate and diastolic blood pressure has been confirmed by a model-based approach \citep{riedl2010short}. However, there was no indication of an interaction between heart rate and systolic blood pressure; only an indirect coupling from heart rate to systolic blood pressure via diastolic blood pressure was found.

\section{Data}
\begin{figure}[hbtp]
\begin{center}
\includegraphics[width=\textwidth]{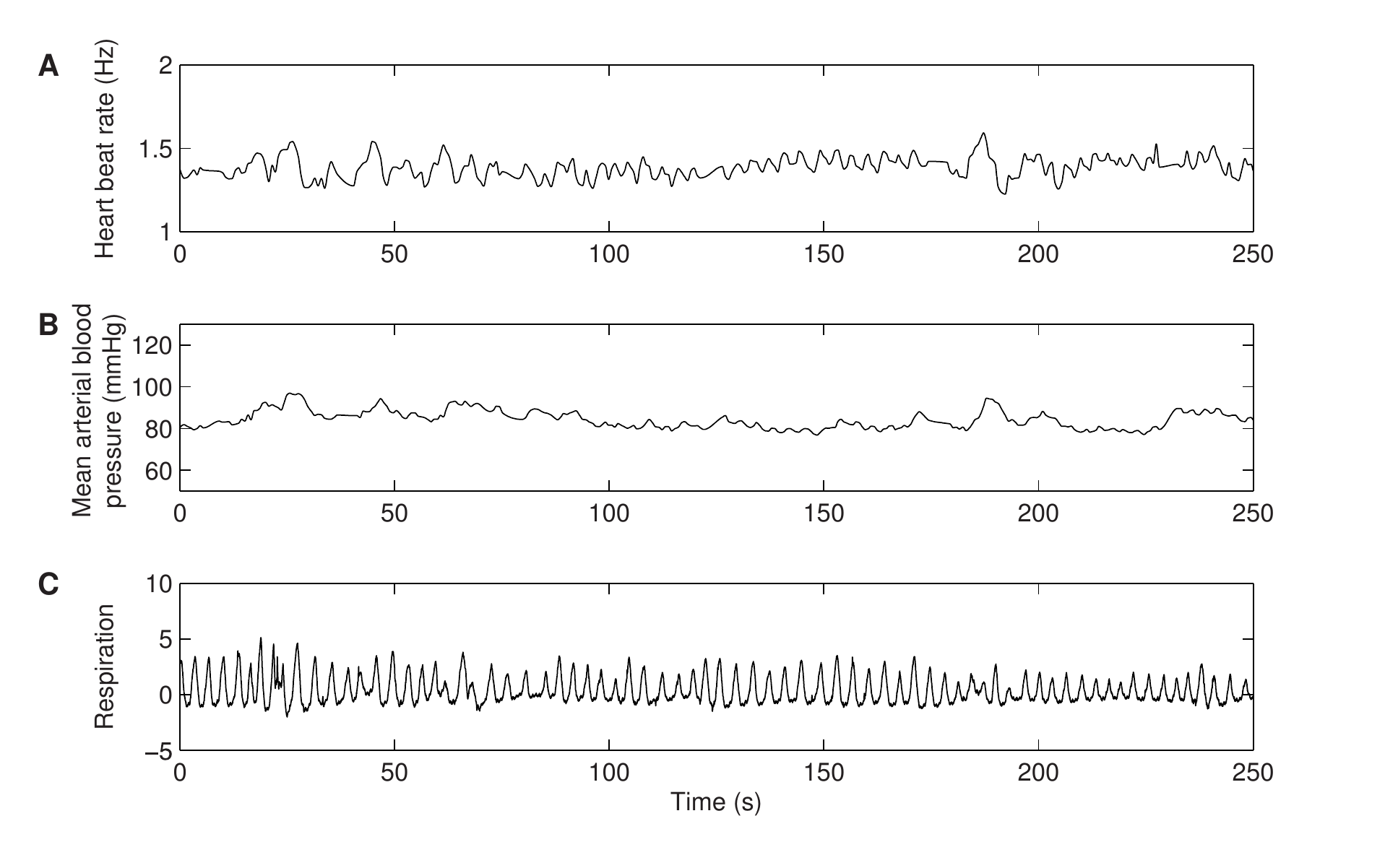}
\caption{Exemplary time series of (A) heart rate, (B) mean arterial blood pressure, and (C) respiration measured on a healthy pregnant woman.}
\label{data}
\end{center}
\end{figure}

Blood pressure and respiration have been measured on eleven pregnant women multiple times (in total 23 data sets) in the course of second and third trimester of pregnancy \citep{malberg2007}. 
The continuous blood pressure was measured non-invasively via finger cuff (100Hz, Portapres device model 2, BMI--TNO, Amsterdam, The Netherlands). 
The respiration curve $R$ was recorded via respiratory effort sensors at the chest (sampling rate 10Hz). 
The measurements were performed for subjects in a supine position with relaxed breathing at times between 8:00h and 12:00h.
Measurements with disturbed respiratory signals or pathological respiratory patterns, e.g., Cheyne-Stokes breathing, have been excluded. 
Based on an algorithm by \citet{suhrbier2006}, we have extracted the heart beats and calculated an average heart beat rate $H$ (in Hz).
The main objective of the analysis of heart rate and blood pressure is to investigate the cardiovascular system during normal sinus rhythm. 
Therefore, we have removed beats not coming from the sinus node of the heart, so-called ventricular premature complexes that are not directly controlled by the autonomous nervous system. 
These features are exchanged for random values by an adaptive filter algorithm preserving the time relation (http://tocsy.pik-potsdam.de; \citep{wessel2000}).
\textcolor{black}{The ratio of the frequency of the respiration and the heart signal does not necessarily have to be an integer number. Therefore, some important variability of the respiration signal is lost if resampled to the beat-to-beat based time-scale. In order to consider the entire variability of the respiration, we have to use the continuous signals.}
Systolic $S$ and diastolic $D$ blood pressures are estimated from the maxima and minima of the blood pressure curve.
Instead using systolic $S$ and diastolic $D$ blood pressures, we have calculated the mean brachial blood pressure $B = D + \frac{1}{3}(S - D)$ \textcolor{black}{and interpolated it to a ``continuous'' signal, because we are interested in temporal continuous values, but both, $D$ and $S$, can only be used in an analysis on a beat-to-beat basis.}
All time series have been resampled to 10Hz (Fig.~\ref{data}). \textcolor{black}{For the average heart beat rate $H$ and the mean brachial blood pressure $B$ this was done by using a cubic interpolation.}

\section{Method}
A recurrence plot is a representation of recurrent states of a dynamical system $X$ in its $m$-dimensional phase space. 
A phase space trajectory can be reconstructed from a time series $\{u_i\}_{i=1}^N$ by time delay embedding \citep{packard80}
\begin{equation}
\vec{x}_i = (u_i, u_{i+\tau}, \ldots, u_{i+\tau (m-1)}),
\end{equation}
where $m$ is the embedding dimension, $\tau$ is the delay, and $N'=N-(m-1)\tau$ is the number of phase space vectors. 

The embedding parameters $\tau$ and $m$ can be estimated by using mutual information and false nearest neighbour method \citep{kantz97}. 

\begin{figure}[hbtp]
\begin{center}
\includegraphics[width=.9\textwidth]{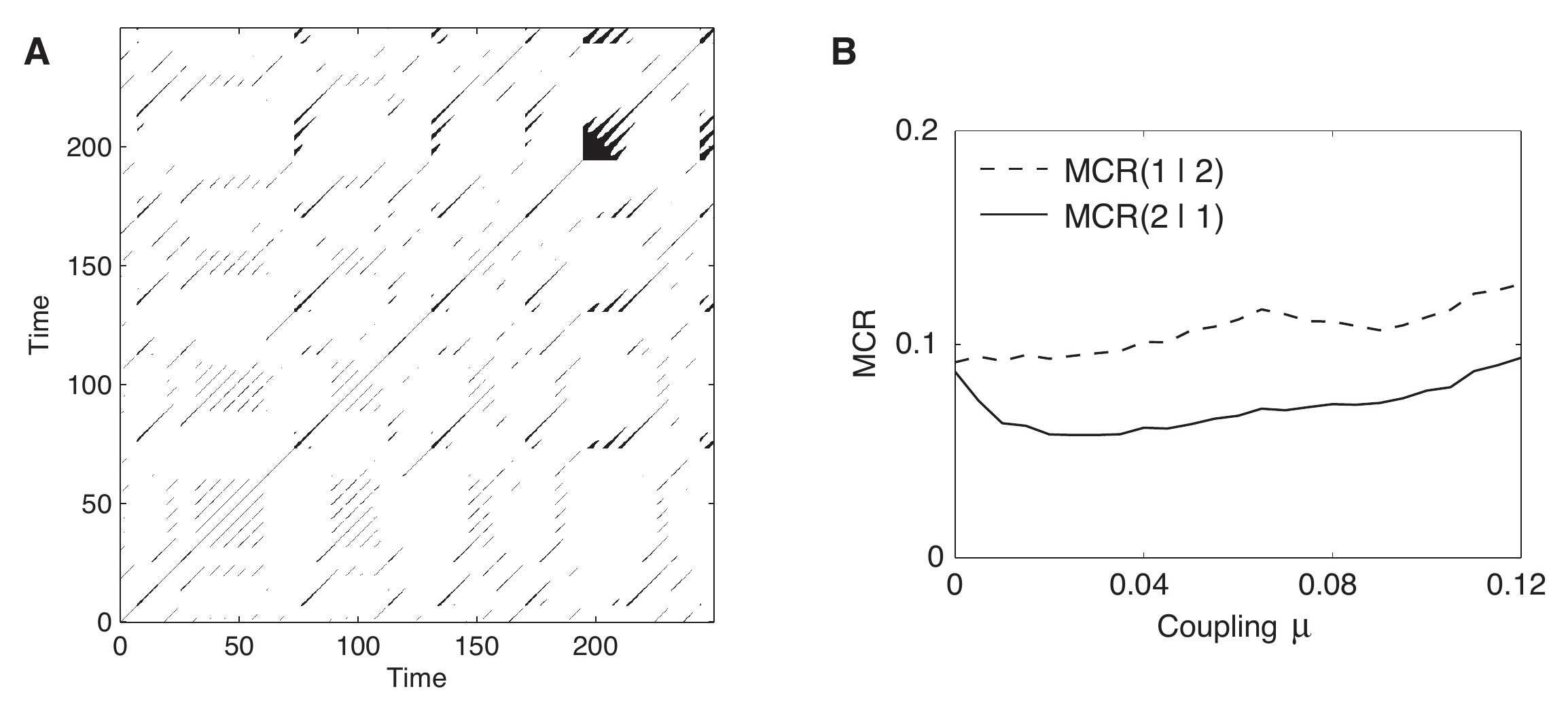}
\caption{(A) Recurrence plot of the R\"ossler system
$(\dot x, \dot y, \dot z) = 
( - y - z,\ x + a \, y,\ b + z \, (x - c))$
with $a=b=0.2$ and $c=10$ \citep{roessler1976}.
(B) Mean conditional probability of recurrence of two coupled R\"ossler systems
$(\dot x_i, \dot y_i, \dot z_i) = 
( - (0.99 + \nu_i) y_i - z_i + \mu_i(x_j - x_i),\ (0.99 + \nu_i) x_i + a \, y_i,\ b + z_i \, (x_i - c))$
with $a=b=0.2$, $c=10$, frequency mismatch $\nu_1 = 0.05$ and $\nu_2 = -0.05$, 
and coupling strength $\mu_1 = 0$ and $\mu_2=[0, \ldots, 0.12]$ between the two R\"ossler systems $i=1, 2$.
System 1 drives system 2 ($\mu_1=0$), therefore, $MCR(2|1) < MCR(1|2)$.}
\label{mcr_roessler}
\end{center}
\end{figure}

The RP is then a pair-wise test of all phase space vectors $\vec{x}_i$ ($i=1,\ldots,N', \vec{x} \in \mathbb{R}^m$) among each other, whether or not they are close:
\begin{equation}
R_{i,j}^{X} = \Theta\bigl(\varepsilon - ||\vec{x}_i - \vec{x}_j||\bigr),
\end{equation}
with $\Theta(\cdot)$ being the Heaviside function, $||\vec{x}_i - \vec{x}_j||$ the spatial distance between the phase space vectors, and $\varepsilon$ a threshold for proximity \citep{marwan2007}. 
The indices $i$ and $j$ range in the interval $[1, \ldots,N']$ and mark the time points along the phase space trajectory of length $N'$.
The binary recurrence matrix $\mathbf{R}^{X}$ contains the value one for all close pairs (Fig.~\ref{mcr_roessler}A). 

The average of the recurrence matrix $\langle p (\vec x)\rangle = \sum_{i,j} R_{i,j}^{X}/N'^2$ is called recurrence rate and corresponds to the mean probability that {\it any} state recurs. 
The probability, that the system recurs to a {\it certain} state $\vec x_j$ can be estimated by the average of the corresponding column of the recurrence matrix, $p (\vec x_j) = \sum_{i} R_{i,j}^{X}/N'$.
For two coupled systems $X$ and $Y$, we may ask for joint probabilities of recurrences in both systems. Such joint probabilities can be estimated from the joint RP,
\begin{equation}
JR_{i,j}^{X,Y} = \Theta\bigl(\varepsilon - ||\vec{x}_i - \vec{x}_j||\bigr) \times \Theta\bigl(\varepsilon - ||\vec{y}_i - \vec{y}_j||\bigr),
\end{equation}
which represents simultaneous recurrences in systems $X$ and $Y$. Analogously to the recurrence rate, averaging the matrix $\mathbf{JR}^{X,Y}$ delivers the joint recurrence rate, i.e., the probability $p( \vec x_j, \vec y_j)$ that we find a recurrence in system $X$ and in system $Y$ simultaneously. Thus, we can calculate the probability that the trajectory of $Y$ recurs to the neighborhood of $\vec y_j$ under the condition that the trajectory of $X$ recurs to the neighborhood of $\vec x_j$ by
\begin{equation}
p(\vec y_j| \vec x_j) = \frac{p( \vec x_j, \vec y_j)}{p (\vec x_j)} =  \frac{ \sum_{i=1}^{N'} JR_{i,j}^{X,Y}}{\sum_{i=1}^{N'} R_{i,j}^{X}}.
\end{equation}
Its average is the {\it mean conditional probability of recurrence, MCR} \citep{romano2007,zou2011}:
\begin{equation}
MCR(Y|X) = \frac{1}{N'} \sum_j p(\vec y_j| \vec x_j)
\quad \text{and} \quad
MCR(X|Y) = \frac{1}{N'} \sum_j p(\vec x_j| \vec y_j).
\end{equation}

In the presence of the asymmetry of the coupling (e.g., suppose $X$ to be driver and $Y$ to be response without the loss of generality), we have the relationship
\begin{equation}\label{mcr_comp}
MCR(Y |X) < MCR(X |Y ).
\end{equation}
The interpretation of this inequality is based on the difference of complexity between $X$ and $Y$. 
If $X$ drives $Y$, the dimension of $Y$ will be larger than the dimension of $X$ because the dynamics of $Y$ is determined by both the states of $X$ and $Y$, while $Y$ does not influence $X$. 
Note that this only holds provided the coupling strength is smaller than the threshold for synchronization, as the coupling direction might be lost if both systems become completely synchronized. 
Increasing the coupling strength from $X$ to $Y$, the complexity of $Y$ increases. 
This results in a decrease of the recurrence probability $p(\vec y_j)$ that $Y$ recurs to the neighborhood. 
However, the complexity of $X$ remains constant with increasing coupling strength because $X$ is independent of $Y$ (not vice versa). 
Hence, the mean recurrence probability of $\langle p(\vec x_j )\rangle > \langle p(\vec y_j )\rangle$, implying $\sum_i R_{i,j}^X > \sum_i R_{i,j}^Y$. 
Therefore, we have $MCR(Y |X) < MCR(X | Y )$ if $X$ is the driver (Fig.~\ref{mcr_roessler}B).

Following \cite{romano2007}, the criterion for selecting the threshold value $\varepsilon_X$ and $\varepsilon_Y$ is such that for coupling
strength equal to zero the recurrence rates (recurrence probabilties) in both systems should be equal. However, in a passive experiment where the coupling strength between both interacting systems cannot be adjusted systematically
like our measurement data, we cannot apply directly such criterion to choose $\varepsilon_X$ and $\varepsilon_Y$, because the value of coupling strength is
not known per se. Therefore, we apply another criterion in
choosing the threshold \citep{schinkel2008}: we normalize the data beforehand to have zero mean
and unit standard deviation, and then we choose
$\varepsilon_X=\varepsilon_Y=0.1$. Consistent results are obtained
for threshold values that are varied in the range [0.05, 0.2]. 

In this
work, we are interested in testing the possible interaction direction between a
pair of two time series. An investigation of indirect couplings between the
three subsystems requires a systematic study involving all three subsystems simultaneously which will be future work.

In the case of passive experiments, we often have one scalar
measurement time series like what we have for the cardiovascular experiment. 
We need to statistically assess the significance of the so calculated (often just one) direction value in order to
decide whether the value is obtained by chance or whether it is significant. 
Therefore, we need an appropriate statistical test in order to test the null hypothesis that
the two systems $X$ and $Y$ have an independent recurrence structure. To test such null hypothesis, we
use the phase randomization surrogate test \citep{prichard1994}. Random phases are added
to the Fourier transformed time series which is then inversely transformed to derive the new
time series (with different phases). When assessing the
$MCR(X|Y)$ value (where $X$ and $Y$ represent $R, H$ and $B$ series, respectively) of one subject, we
use the phase randomized time series of the second time series $Y$
as a surrogate $Y^s$ (we can also use the first time series $X$ to create
surrogates $X^s$, but it does not change the results). Repeating the phase randomization (in our work 100 times),
we get an ensemble of many surrogate series $Y^s$ and, hence, a distribution of
corresponding $MCR$ values.
The directionality indices $MCR(X|Y)$ and $MCR(Y|X)$ for one subject can now
be compared with the distribution of $MCR(X|Y^s)$ and
$MCR(Y^s|X)$, respectively. If $X$ and $Y$ are independent, the value $MCR(X|Y)$ will
not differ significantly from the distribution of the values $MCR(X|Y^s)$.
Otherwise, i.e., when exceeding the 0.95-quantile, we can reject the null hypothesis, indicating that the obtained
values for the directionality indices are significant with 95\% confidence. 

Summarizing, the
following steps have to be undertaken to assess the coupling direction between
two time series for each subject:
\begin{enumerate}
\item[(i)] Choose the significance level $\alpha=0.05$. 
\item[(ii)] Compute $MCR(X|Y)$ and $MCR(Y|X)$. 
\item[(iii)] Create 100 phase randomized surrogates and compute $MCR(X|Y^{s_j})$ and $MCR(Y^{s_j} | X)$ for $j=1, \dots,
100$. 
\item[(iv)] Calculate the $\alpha$-quantiles of the distributions of $MCR(X|Y^{s})$ and $MCR(Y^{s} | X)$.
\item[(v)] If $MCR(X|Y)$ and $MCR(Y|X)$ are larger than the corresponding
$\alpha$-quantiles, reject the null hypothesis and consider them as significant.
\item[(vi)] If $MCR(X|Y)$ and $MCR(Y|X)$ are significant, we
compare $MCR(X|Y)$ and $MCR(Y|X)$ regarding Eq.~(\ref{mcr_comp}) in order to find the directionality of the coupling.
\end{enumerate}

\section{Results}

Using mutual information and the method of false nearest neighbours, we have found optimal embedding parameters for $H$ as well as for $R$ to be $\tau=2$ and $m=3$, which resulted from the average over all cases; for $B$ we have found $\tau=4$ and $m=2$. The results of our analysis have not changed much when using different embedding parameters.

\begin{figure}[hbtp]
\begin{center}
\includegraphics[width=\textwidth]{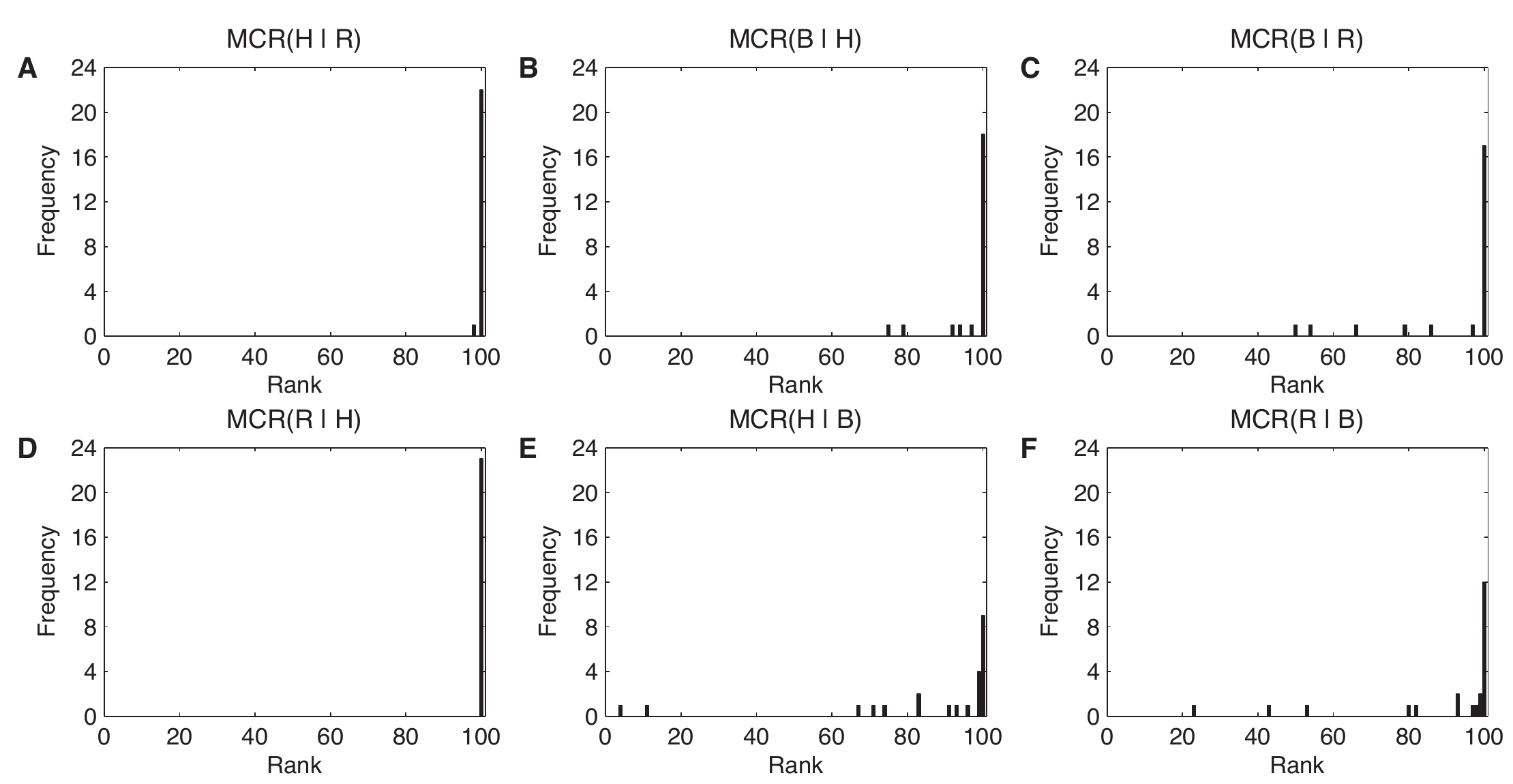}
\caption{Significance test using phase-randomized surrogates for 
(A) $MCR(H|R)$, (B) $MCR(B|H)$, (C) $MCR(B|R)$, (D) $MCR(R|H)$, (E) $MCR(H|B)$,
(F) $MCR(R|B)$.}
\label{significancetest}
\end{center}
\end{figure}

We have calculated the $MCR$ measures for all combinations between respiration $R$ and heart rate $H$, heart rate $H$ and mean blood pressure $B$, and respiration $R$ and mean blood pressure $B$.
First we check the significance of $MCR$ in order to limit the subsequent directionality study to the significant results. A $MCR$ value would be significant if it exceeds the 0.95-quantile of the surrogate $MCR$ distribution. Based on this test, 
we find significant MCR indices between respiration $R$ and heart rate $H$ for all subjects, but between mean blood pressure and heart rate or respiration only for more than half of the subjects, although still a considerable number
(Fig.~\ref{significancetest}, Tab.~\ref{tabsignificancetest}).

\begin{table}[htbp]
\caption{Number of significant $MCR$ values (extending the  0.95-quantile of the test distribution).}
\label{tabsignificancetest}
\centering \begin{tabular}{lr}
\hline
Coupling	&Number\\
\hline
$MCR(H|R)$	&23\\
$MCR(R|H)$	&23\\
$MCR(H|B)$	&19\\
$MCR(B|H)$	&14\\
$MCR(B|R)$	&18\\
$MCR(R|B)$	&16\\
\hline
\end{tabular}
\end{table}

\begin{figure}[hbtp]
\begin{center}
\includegraphics[width=\textwidth]{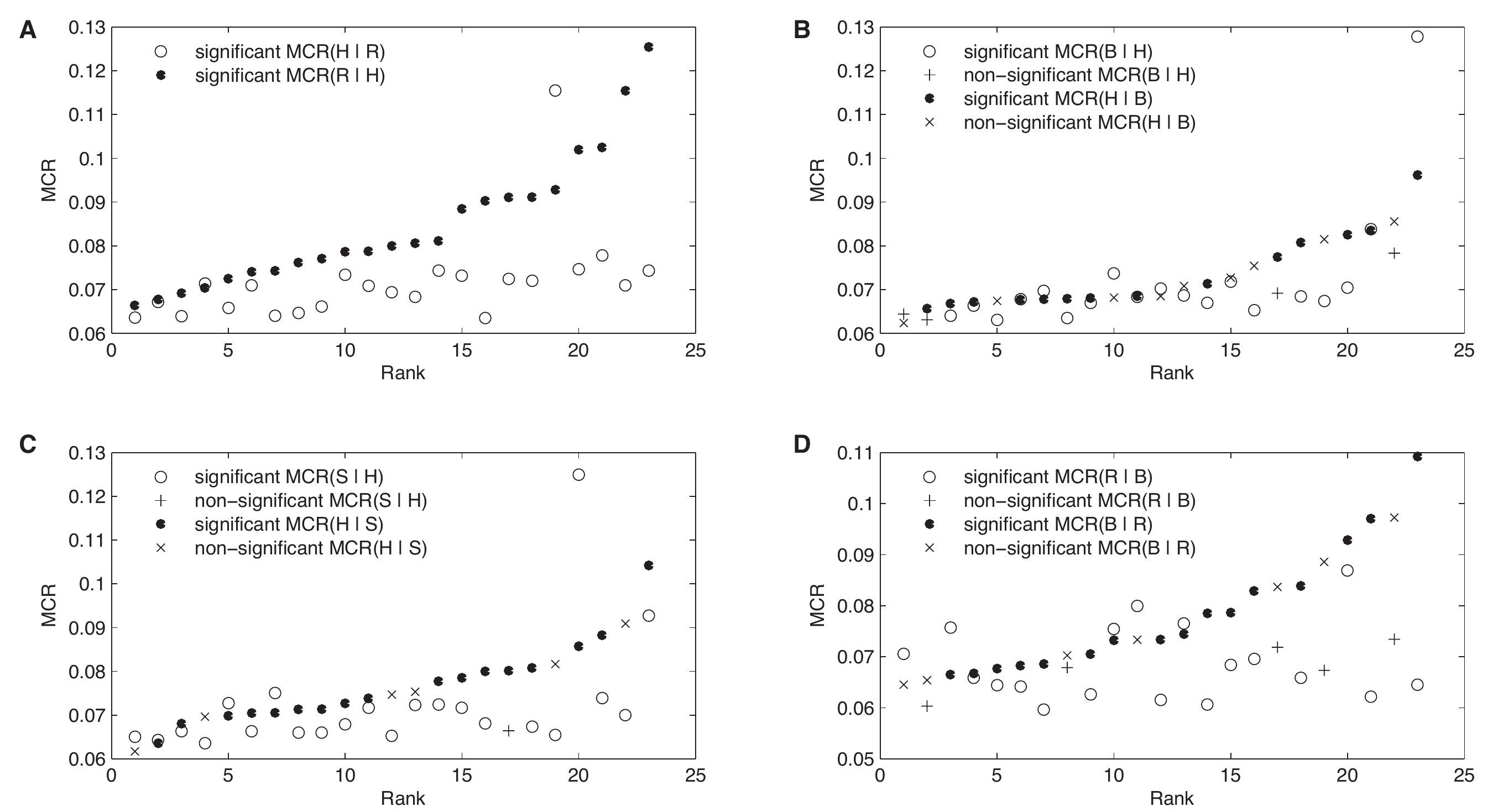}
\caption{(A) $MCR(H|R)$ values together with the sorted significant values of $MCR(R|H)$. In general, $MCR(R|H)$
is larger than $MCR(H|R)$, indicating a coupling direction from respiration to heart rate.
(B) $MCR(B|H)$ values together with the sorted significant and non-significant values of $MCR(H|B)$.
Larger $MCR(H|B)$ than $MCR(B|H)$ indicates a coupling from heart rate towards blood pressure.
(C) The same as for (B) but using the upper envelope of the blood pressure $S$ instead of mean blood pressure $B$.
(D) The same as for (A) for mean blood pressure $B$ and respiration $R$. Here the MCR
indices reveal opposite coupling directions from $R$ to $B$ (in 5 subjects) and from $B$ to $R$ 
(in 13 subjects).
}
\label{mcr_cardio}
\end{center}
\end{figure}

Next, we study the coupling direction between the significant couplings. According Eq.~\ref{mcr_comp}, 
we compare which MCR value is larger. First, we check the coupling direction between
respiration $R$ and heart rate $H$ (Fig.~\ref{mcr_cardio}A). We find 21 significant cases where
the $MCR(R|H)$ value is clearly larger than $MCR(H|R)$, thus, we can infer a coupling direction 
from respiration to heart rate. Preeclampsia and the progression of gestation have 
not caused the significance of $MCR(H|R)$.

Then, we check the coupling between heart rate $H$ and blood pressure $B$ (Fig.~\ref{mcr_cardio}B).
Here we find 15 significant cases where $MCR(H|B)$ is larger than $MCR(B|H)$, i.e., 
a coupling from heart rate $H$ to blood pressure $B$. Including the nonsignificant $MCR(H|B)$ values, we 
would even have 18 cases with such coupling direction. In 5 cases, we found an opposite 
coupling direction from blood pressure towards hear rate. However, the difference
between the two MCR indices is in more than 15 cases small, indicating a potential 
bidirectional coupling.

Finally, the comparison between the significant MCR values of blood pressure $B$ and respiration $R$
reveals 13 cases with coupling directions from blood pressure to respiration and 5 cases from 
respiration to blood pressure (Fig.~\ref{mcr_cardio}D).

The coupling directions between heart rate and blood pressure as well as between respiration and blood pressure 
are not as clear as between respiration and heart rate, because the differences of the
corresponding two MCR measures is small (Fig.~\ref{mcr_cardio}B, D) and there are also some cases with
opposite coupling directions (e.g., where $MCR(B|H)$ is larger than $MCR(H|B)$ in Fig.~\ref{mcr_cardio}B). 
We might guess that this latter result could be due to preeclampsia (PE).
However, this is not the case, as the contradictory results appear for preeclampsia as well as for 
healthy women (for $H$ vs.~$B$ in 2 healthy and 1 PE, for
$R$ vs.~$B$ in 3 healthy and 1 PE women).

Instead using the mean blood pressure $B$, we have also tested the upper envelope of the blood 
pressure series $S$ (as an analog for a continuous systolic blood pressure).
This upper envelope can be interpreted as a representation of the current total peripheral resistance of the smaller 
blood vessels. Here we found larger differences between $MCR(H|B)$ and $MCR(B|H)$ and finally
18 significant cases with a coupling from heart rate $H$ to blood pressure $B$
(Fig.~\ref{mcr_cardio}C).
This might be indicative for the mechanical coupling mechanisms affecting the blood pressure by the
heart rate \citep{mullen1997system}.

\section{Conclusions}
The investigation of coupling directions from experimental data is a challenging task \citep{palus2007}. 
The recently developed non-linear method based on conditional recurrence probabilities \citep{romano2007}
allows for a directionality analysis in coupled complex systems. Here we have successfully applied this approach for a coupling analysis in experimental data. 

The application on data from the human cardio-respiratory system
has clearly revealed a coupling from the respiratory system towards the heart. 
These findings support the assumption that the respiratory sinus-arrhythmia results from a direct influence of respiration on heart rate (respiratory gate \citep{eckberg2003}). It is assumed that the respiratory control centres modulate the vagal outflow in the brainstem.


Cardiovascular couplings are not as clearly detected than the respiratory coupling, which suggest
that the respiratory induced oscillation is the carrier of the couplings detected by the beat-to-beat approaches \citep{porta2011causal,suhrbier2010cardiovascular,faes2010extended,baumert2002joint,riedl2010short,faes2012non}. 
The proposed method has been able to detect that heart rate affects blood pressure (through mechanical
coupling mechanisms \citep{taylor1996fundamental,mullen1997system}), but for some cases 
also a that arterial pressure affects heart rate (through the baroflex circuit). The small difference
between the MCR measures also supports the potential bidirectional nature of the coupling between heart
rate and blood pressure. An even less clear result
was found for the coupling between respiration and blood pressure. This might be a hint 
to indirect coupling mechanisms. Moreover, here we have used continuous cardio-respiratory
signals instead of beat-to-beat based signals, which was the base in previous studies. 
In contrast to beat-to-beat signals,
in the blood pressure series the high frequency variation is suppressed. 
These distinctions \textcolor{black}{and also the fact, that we extracted the heart rate from the blood pressure measurements,} might cause the differences in the found coupling structure.
Nevertheless, the proposed method could lead to additional information about the cardiorespiratory coupling in comparison to the beat-to-beat approach. 

The particular database used in our study might also have some impact on our findings. 
Nevertheless, during our analysis we had not found any evidence that either preeclampsia or the progression of gestation had a significant impact on the results.
However, a detailed analysis of the
specific effect of pregnancy and preeclampsia on the cardio-respiratory coupling is out
of the scope of this paper but is subject to future studies.
\textcolor{black}{Moreover, a thorough study about the accuracy of the detection of interaction directions (e.g., how much should $MCR(X|Y)$ differ from $MCR(Y|X)$) is, in general, an open problem and also remains a subject for future work.}



\subsection{Acknowledgements}
This study was supported by the Potsdam Research Cluster for Georisk Analysis, Environmental Change and Sustainability (PROGRESS, support code 03IS2191B), the DFG research group ``Himalaya: Modern and Past Climates (HIMPAC)'', the Hong Kong Polytechnic University Postdoctoral Fellowship, and the National Natural Science Foundation of China (Grant No.~11135001).


\end{document}